\documentstyle[]{mn}
\newif\ifAMStwofonts

\def\etal{{\rm et al.}}

\def\simgt{\mathrel{\spose{\lower 3pt\hbox{$\sim$}}
        \raise 2.0pt\hbox{$>$}}}
\def\simlt{\mathrel{\spose{\lower 3pt\hbox{$\sim$}}
        \raise 2.0pt\hbox{$<$}}}

\ifoldfss
  \ifCUPmtlplainloaded \else
    \NewTextAlphabet{textbfit} {cmbxti10} {}
    \NewTextAlphabet{textbfss} {cmssbx10} {}
    \NewMathAlphabet{mathbfit} {cmbxti10} {} 
    \NewMathAlphabet{mathbfss} {cmssbx10} {} 
  \fi
  \ifAMStwofonts
    \ifCUPmtlplainloaded \else
      \NewSymbolFont{upmath} {eurm10}
      \NewSymbolFont{AMSa} {msam10}
      \NewMathSymbol{\upi}     {0}{upmath}{19}
      \NewMathSymbol{\umu}     {0}{upmath}{16}
      \NewMathSymbol{\upartial}{0}{upmath}{40}
      \NewMathSymbol{\leqslant}{3}{AMSa}{36}
      \NewMathSymbol{\geqslant}{3}{AMSa}{3E}

    \fi
  \fi
\fi 

\ifnfssone
  \newmathalphabet{\mathit}
  \addtoversion{normal}{\mathit}{cmr}{m}{it}
  \addtoversion{bold}{\mathit}{cmr}{bx}{it}
  \newmathalphabet{\mathbfit} 
  \addtoversion{normal}{\mathbfit}{cmr}{bx}{it}
  \addtoversion{bold}{\mathbfit}{cmr}{bx}{it}
  \newmathalphabet{\mathbfss} 
  \addtoversion{normal}{\mathbfss}{cmss}{bx}{n}
  \addtoversion{bold}{\mathbfss}{cmss}{bx}{n}
  \ifAMStwofonts
    \ifCUPmtlplainloaded \else
      \UseAMStwoboldmath
      \makeatletter
      \new@mathgroup\upmath@group
      \define@mathgroup\mv@normal\upmath@group{eur}{m}{n}
      \define@mathgroup\mv@bold\upmath@group{eur}{b}{n}
      \edef\UPM{\hexnumber\upmath@group}
      \new@mathgroup\amsa@group
      \define@mathgroup\mv@normal\amsa@group{msa}{m}{n}
      \define@mathgroup\mv@bold\amsa@group{msa}{m}{n}
      \edef\AMSa{\hexnumber\amsa@group}
      \makeatother
      \mathchardef\upi="0\UPM19
      \mathchardef\umu="0\UPM16
      \mathchardef\upartial="0\UPM40
      \mathchardef\leqslant="3\AMSa36
      \mathchardef\geqslant="3\AMSa3E
    \fi
  \fi
\fi 

\ifnfsstwo
  \DeclareMathAlphabet{\mathbfit}{OT1}{cmr}{bx}{it}
  \SetMathAlphabet\mathbfit{bold}{OT1}{cmr}{bx}{it}
  \DeclareMathAlphabet{\mathbfss}{OT1}{cmss}{bx}{n}
  \SetMathAlphabet\mathbfss{bold}{OT1}{cmss}{bx}{n}
  \ifAMStwofonts
    \ifCUPmtlplainloaded \else
      \DeclareSymbolFont{UPM}{U}{eur}{m}{n}
      \SetSymbolFont{UPM}{bold}{U}{eur}{b}{n}
      \DeclareSymbolFont{AMSa}{U}{msa}{m}{n}
      \DeclareMathSymbol{\upi}{0}{UPM}{"19}
      \DeclareMathSymbol{\umu}{0}{UPM}{"16}
      \DeclareMathSymbol{\upartial}{0}{UPM}{"40}
      \DeclareMathSymbol{\leqslant}{3}{AMSa}{"36}
      \DeclareMathSymbol{\geqslant}{3}{AMSa}{"3E}
    \fi
  \fi
\fi 

\ifCUPmtlplainloaded \else
  \ifAMStwofonts \else 
    \def\upi{\pi}
    \def\umu{\mu}
    \def\upartial{\partial}
  \fi
\fi

\title[Determining the Microlens Mass Function from Quasar Microlensing Statistics]
  {Determining the Microlens Mass Function from Quasar Microlensing Statistics}
\author[Wyithe \& Turner]
  {J.~S.~B.~Wyithe$^{1,2}$, 
  E.~L.~Turner$^2$ \\
  $^1$ School of Physics, The University of Melbourne, Parkville, Vic, 3052, 
Australia\\
  $^2$ Princeton University Observatory, Peyton Hall, Princeton, NJ 08544, USA\\ 
 Email: swyithe@astro.Princeton.edu, elt@astro.Princeton.edu }
\date{Accepted. Received}
\pagerange{\pageref{firstpage}--\pageref{lastpage}}
\pubyear{1999}

\def\LaTeX{L\kern-.36em\raise.3ex\hbox{a}\kern-.15em
    T\kern-.1667em\lower.7ex\hbox{E}\kern-.125emX}

\begin{document}

\label{firstpage}

\maketitle

\begin{abstract}

The first investigations of the response of the microlensing magnification pattern (at an optical depth of order unity) to the mass function of the microlenses found that the resulting statistics depend only on the mean microlens mass $\langle m\rangle$. In particular the mean microlensing caustic crossing rate was found to be proportional to $\sqrt{\langle m\rangle}$. We show that while this is true in the limit of mass functions with a narrow range of mass, in general the magnification pattern shows structure that reflects the contribution to the optical depth of microlenses with different masses. We present a better approximation, relating the microlens mass function to light-curve statistics. We show that the variability statistics of quasar microlensing light-curves can (in principle) be inverted to obtain the mass function of the microlenses in the mass range over which the mass density remains comparable, ie. $p(m)dm\approx Cm^{-1}$. A preliminary analysis of the structure function for Q2237+0305 suggests that there is not a significant contribution to the optical depth from very low mass objects ($10^{-3}M_{\odot}$). However observations of multiple microlensed quasars for a period of $\sim 20$ years may in the future yield a detailed $p(m)dm$. In the mass range where the number density is comparable, ie. $p(m)dm\approx const.$, the distribution of flux factors could be inverted to find the microlens mass function. This may be used as a probe of the abundance of planets with orbital radii $>100$AU. 

\end{abstract}
 

\section{Introduction}

Following the findings of Witt, Kayser \& Refsdal (1993) and Lewis \& Irwin (1995, 1996) the mean microlens mass has been assumed to be the only important mass function parameter with respect to quasar microlensing statistics. The primary implication of this assumption for the determination of the microlens mass function is that observations of quasar microlensing could yield a value for the microlens mass independently from the detail of the mass function. However the converse, that quasar microlensing provides no information about the detail of the microlens mass function (Witt, Kayser \& Refsdal 1993) is contrary to the initial hopes that such information could be extracted from long term monitoring of objects like Q2237+0305 (e.g. Wambsganss, Paczynski \& Schneider 1990). 

Lewis \& Irwin (1995) showed that the point source magnification distribution is independent of the mass function. However for a given set of macro-parameters they also show that the magnification distribution is affected by a component of optical depth in smooth matter ($\kappa_c$). On the other hand it is not surprising that the magnification distribution depends on the optical depth ($\kappa$) and the shear ($\gamma$). The dependence on $\kappa_c$ is therefore readily understood in light of the parameter transformation of Paczynski (1986) which relates models containing continuous matter to those that include only point masses but are described by different macro parameters. In combination with a macrolens model, the magnification distribution provides a means to probe the fraction of continuous matter (Lewis \& Irwin 1995). Of course the effect of a finite source size, which narrows the width of the magnification distribution, also needs to be considered.

Refsdal \& Stabell (1993) noted qualitatively that if both small and large masses contribute significantly to the optical depth, one expects that, for a source size lying between the respective microlens Einstein Radii (ER), small fluctuations will be superimposed on the large scale variations caused by microlensing due to the larger masses. This statement must be true for the following reason. If the example is taken to its extreme, by decreasing the small mass and at the same time keeping the optical depth due to the small mass population constant (ie. by increasing their number density), then the model acts like one that includes point masses plus a continuous sheet. The mean mass is zero, but the microlensing rate certainly is not. Clearly microlensing variability is not determined solely by $\sqrt{\langle m\rangle}$.

In this paper we discuss the effect of a mass function on various microlensing statistics, and show that in principle the microlens mass function can be retrieved from long-term monitoring of microlensed quasars. Sec. \ref{magmaps} presents magnification patterns for different mass functions, and gives a qualitative discussion of their features. In Sec. \ref{invert} structure functions and flux factor distributions are shown. Approximate expressions are presented which relate the microlens mass function to the microlensing light-curve statistics. Finally in Secs. \ref{2237app} and \ref{discussion} we present a preliminary analysis of the structure function for Q2237+0305, and discuss potential applications.

\section{Magnification Patterns}
\label{magmaps}

\begin{figure*}
\vspace{200mm}
\includegraphics{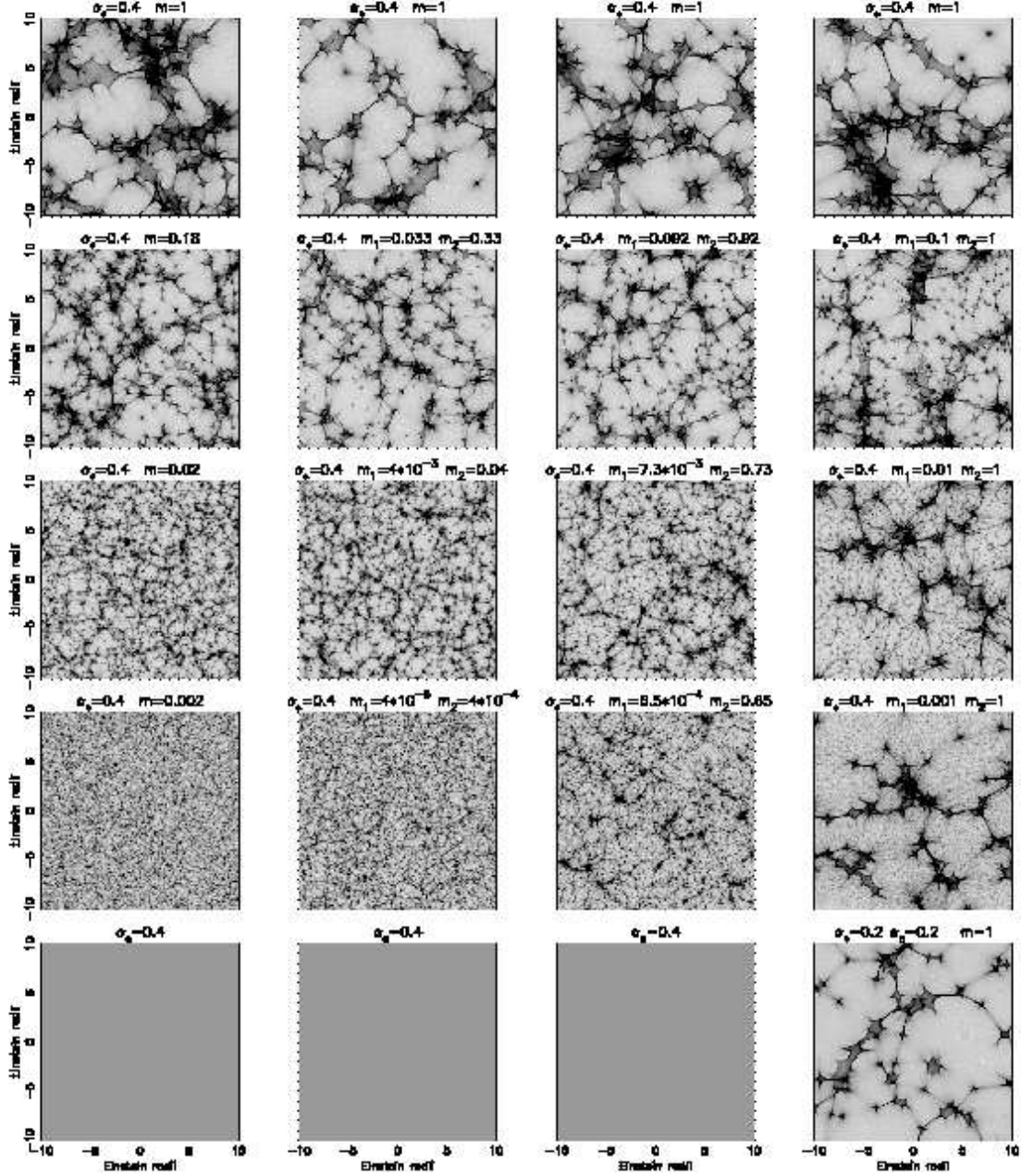}
\caption{\label{magmap1} Magnification patterns for different mass functions. Each row has a different mean microlens mass (shown in column 1), and each of the columns 1-4 a different mass spectrum (mass functions $p_{Single}$, $p_{Ndense}$, $p_{Salpeter}$ and $p_{Odepth}$ respectively). The upper and lower mass limits are shown by $m_1$ and $m_2$ in each case. The bottom row shows the limiting case where $\frac{m_1}{m_2}\rightarrow 0$.}
\end{figure*}

\begin{figure*}
\vspace{160mm}
\includegraphics{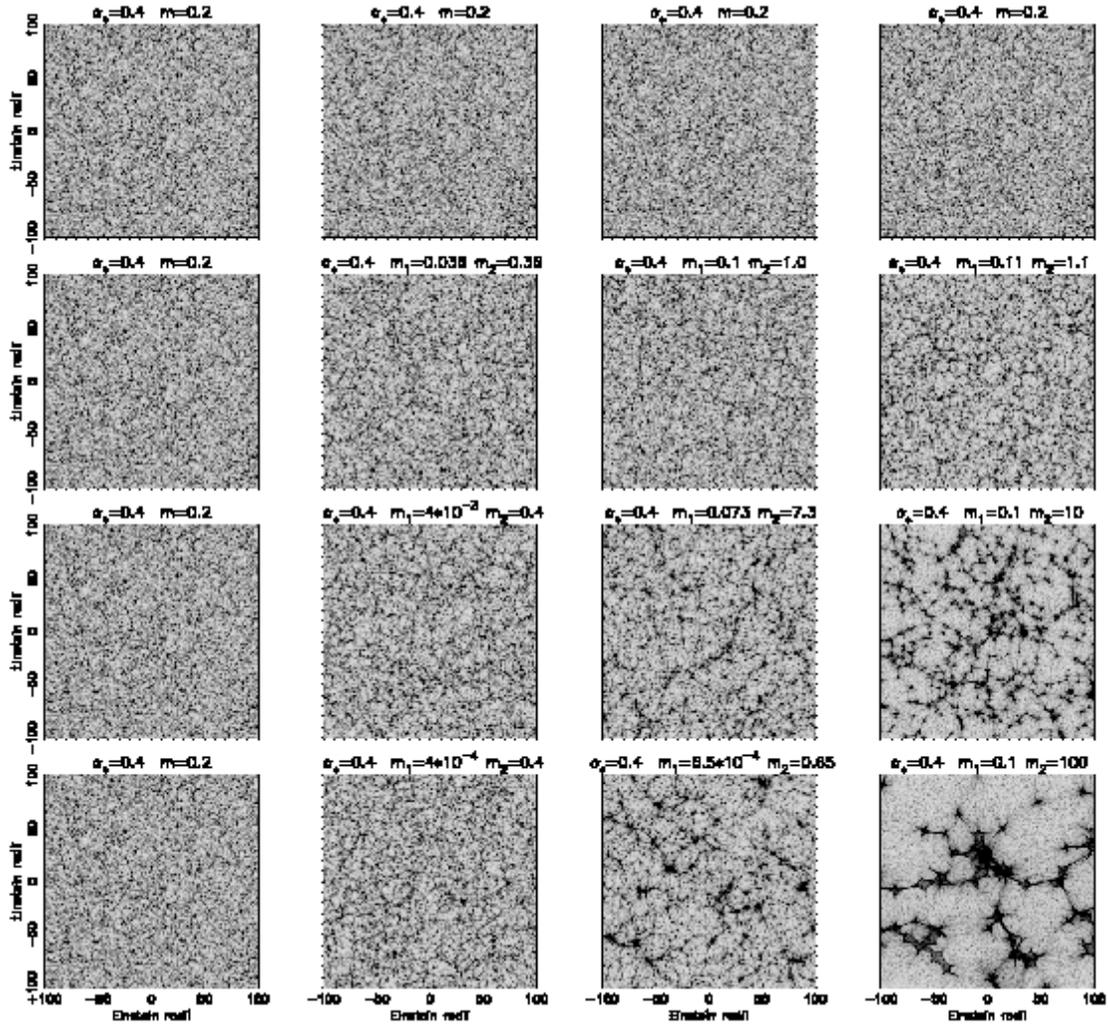}
\caption{\label{magmap2} Magnification patterns for different mass functions, with a mean microlens mass of $\langle m\rangle=0.2M_{\odot}$ in all cases. The columns 1-4 show results from different mass spectra (mass functions $p_{Single}$, $p_{Ndense}$, $p_{Salpeter}$ and $p_{Odepth}$ respectively). The upper and lower mass limits are shown by $m_1$ and $m_2$ in each case.}
\end{figure*}

\begin{figure*}
\vspace{60mm}
\includegraphics{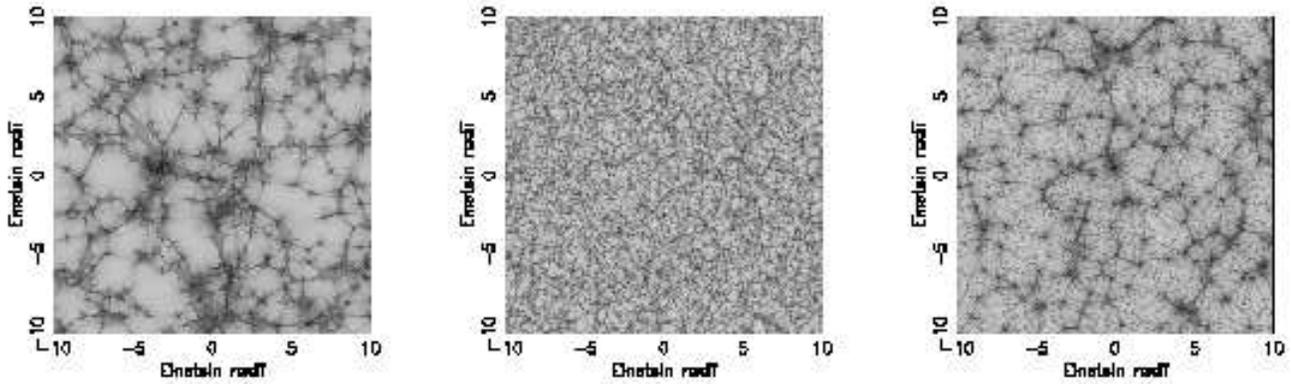}
\caption{\label{superpos} Magnification patterns for single mass microlens populations ($p_{Single}$) of $m=1M_{\odot}$ (left) and $m=0.01M_{\odot}$ (centre), and for a model with $p_{Odepth}$ having $\frac{m_{2}}{m_{1}}=100$ and $m_{2}=1$ (right). The right-hand panel is approximately an average of the left-hand and centre panels.}
\end{figure*}

\begin{table*}
\begin{center}
\caption{\label{tab1}Values for the number of stars used and the average magnification obtained from the simulations presented in Fig. \ref{magmap1}. The theoretical mean is $\mu_{av}=2.78$.}
\begin{tabular}{|c|c|c|c|c|c|}
\hline
                                 &                         &  \multicolumn{4}{c}{Mass function}  \\
$\langle m\rangle$ ($M_{\odot}$) &   $\frac{m_{2}}{m_1}$   &  $p_{Single}$       &  $p_{Ndense}$   &  $p_{Salpeter}$  & $p_{Odepth}$    \\\hline
1.0                              &        1                &592 (3.24)  &592 (2.34)  &592 (2.93)   &592 (2.99)\\     
0.18                             &        10               &1970 (2.91) &2222 (2.44) &2165 (2.56)  &2637 (2.82)  \\
0.02                             &        100              &13308 (2.74)&14193 (2.80)&15643 (2.81) &23576 (2.78) \\
0.002                            &        1000             &117599 (2.78) &120284 (2.79)&136754 (2.79) &233018 (2.85) \\
NA                               &        NA               & NA (2.79)  & NA (2.79)  &  NA (2.79)  & 232 (2.24)  \\\hline
\end{tabular}
\end{center}
\end{table*}

\begin{table*}
\begin{center}
\caption{\label{tab2}Values for the number of stars used and the average magnification obtained from the simulations presented in Fig. \ref{magmap2}. The theoretical mean is $\mu_{av}=2.78$.}
\begin{tabular}{|c|c|c|c|c|c|}
\hline
                                 &                         &  \multicolumn{4}{c}{Mass function}  \\
$\langle m\rangle$ ($M_{\odot}$) &   $\frac{m_{2}}{m_1}$   &  $p_{Single}$       &  $p_{Ndense}$   &  $p_{Salpeter}$  & $p_{Odepth}$    \\\hline
0.2                              &        1                &117599 (2.77) &117599 (2.77) &117599 (2.77)  &117599 (2.77)\\      
0.2                              &        10               &117599 (2.77) &119495 (2.78) &119084 (2.77)  &122428 (2.72)\\
0.2                              &        100              &117599 (2.77) &120202 (2.79) &124354 (2.79)  &145246 (2.82)\\
0.2                              &        1000             &117599 (2.77) &120284 (2.79) &136754 (2.85)  &233018 (2.79)\\\hline
\end{tabular}
\end{center}
\end{table*}

In this section we present magnification patterns produced by microlens populations with different mass functions having upper ($m_{2}$) and lower ($m_{1}$) limits that produce mass ranges of $\frac{m_{2}}{m_{1}}=1$, $\frac{m_{2}}{m_{1}}=10$, $\frac{m_{2}}{m_{1}}=100$ and $\frac{m_{2}}{m_{1}}=1000$. Four classes of mass function have been chosen for illustration:

\begin{equation}
i)\hspace{3mm}p_{Single}(m)dm=\delta(m)
\end{equation}

\begin{equation}
ii)\hspace{3mm}p_{Ndense}(m)dm=\frac{ \delta(m_{1})+\delta(m_{2})}{2}
\end{equation}

\begin{eqnarray}
\nonumber
iii)\hspace{3mm}p_{Salpeter}(m)dm&\propto& m^{-2.35} \hspace{3mm} m_{1}<m<m_{2} \\
\hspace{8mm}&=& 0 \hspace{5mm} {\rm otherwise}
\end{eqnarray}

\begin{equation}
iv)\hspace{3mm}p_{Odepth}(m)dm\propto\frac{m_{2}}{m_{1}}\delta(m_{1})+\delta(m_{2})
\end{equation}

\noindent $p_{single}(m)$ describes a population of identical point masses. $p_{Ndense}(m)$ and $p_{Odepth}(m)$ describe mass distributions with equal number densities and optical depths respectively in the two populations of masses $m_{1}$ and $m_{2}$. $p_{Salpeter}(m)$ is a Salpeter mass function.  

The ray-tracing method (e.g. Kayser, Refsdal \& Stabell 1986; Schneider \& Weiss 1987; Wambsganss, Paczynski \& Katz 1989) was used to calculate the magnification patterns. The number of stars used in the models was calculated through the method described in Lewis \& Irwin (1995) and Wyithe \& Webster (1999). Fig. \ref{magmap1} shows 4 columns of magnification patterns, each 20 ER$_{1M_{\odot}}$ (Einstein radius for 1$M_{\odot}$) on a side. The columns show magnification patterns corresponding to mass functions $p_{Single}(m)$, $p_{Ndense}(m)$, $p_{Salpeter}(m)$ and $p_{Odepth}(m)$, each with an optical depth of $\kappa=0.4$ and no applied shear. The upper 4 rows show magnification maps produced by microlens populations with decreasing $\langle m\rangle$ and increasing mass range ($\frac{m_{2}}{m_{1}}$). The mean is constant in each row and was determined so that $m_{2}=1M_{\odot}$ in column 4 ($p_{Odepth}(m)$):
\begin{equation}
\langle m\rangle = \frac{2}{\frac{m_{2}}{m_{1}}+1}.
\end{equation} 
The bottom row shows the limiting case as $m_{1}\rightarrow0$. The number of stars and the mean magnification respectively for these magnification maps are shown in Tab. \ref{tab1}. 

Columns 2$-$4 of Fig.~\ref{magmap1} demonstrate that the introduction of different scales into the structure of the magnification map results from the inclusion of comparable mass-density from microlenses at opposite ends of the mass range. Fig \ref{magmap1} shows several additional trends. The magnification maps in column 1 ($p_{Single}(m)$) show the well documented increase in uniformity as $m$ is lowered (e.g. Wambsganss, Paczynski \& Katz 1989). The magnification pattern approaches a constant for a small but finite source. This is also true for the other mass functions discussed. However, the magnification maps corresponding to $p_{Ndense}(m)$, $p_{Salpeter}(m)$ and $p_{Odepth}(m)$ in columns 2, 3 and 4 demonstrate a slowing of this trend if mass density is distributed over a finite mass range. For $p_{Odepth}(m)$ where half the mass density is in 1$M_{\odot}$ stars and the other half in $10^{-3} M_{\odot}$ microlenses, there is no increase in uniformity over the single mass case, even though the area in square ER (of the mean mass) is increased by a factor of 500. Rows 3 and 4 ($\frac{m_{2}}{m_{1}}=100,1000$) show a decrease in uniformity of the magnification map were contributions to the mass density of microlenses are more evenly spread across the mass range. Only a small section of magnification map is presented (in Fig. \ref{magmap1}) for the smaller mass range examples. Magnification maps corresponding to the upper 4 rows of Fig. \ref{magmap1} have therefore also been plotted in Fig. \ref{magmap2}, this time with a mean microlens mass of $\langle m\rangle =0.2M_{\odot}$ in all cases (the simulation details are summarised in Tab. \ref{tab2}). Fig. \ref{magmap2} shows the trends already mentioned and in addition demonstrates the more significant loss of uniformity where the mass range is greater. However, note that the magnification map for $p_{Odepth}(m)$ clearly shows larger scale structure than the  magnification map for $p_{Single}(m)$ even when the mass range is only 10 in the former case.

$p_{Odepth}(m)$ may not be physically realistic, however the resulting magnification pattern demonstrates the major point of this paper. The magnification maps produced by microlens mass functions of the form $p_{Odepth}(m)$ appear to be averages of the magnification maps produced by $p_{Single}(m_{2})$ and $p_{Single}(m_{1})$. This is demonstrated by Fig. \ref{superpos}, and suggests that microlenses contribute to the magnification pattern in proportion to their mass density. We explore this statement in the next section.

\section{Light-curve Statistics}
\label{invert}

We investigate the relationship between the mass function and the magnification pattern using two light-curve statistics. The first is the structure function $S(\Delta y)$, defined by: 
\begin{equation}
S(\Delta y)=\langle |M(y)-M(y+\Delta y)|\rangle,
\end{equation}
where $y$ is the source position (or equivalently an epoch), and $M(y)$ is the corresponding image magnitude. The second statistic is the distribution of flux factors $p(K)dK$ where $K$ is defined by the near caustic approximation of Chang \& Refsdal (1979):  
\begin{equation}
\label{nearcaust}
\mu = \mu_o + \Theta(y)\frac{K}{\sqrt{\Delta y_{caust}}}.
\end{equation}
In Eqn. \ref{nearcaust}, $\mu$ is the point source magnification, $\mu_o$ is the magnification of non-critical images, $\Delta y_{caust}$ is the distance of the source from the caustic, and $\Theta$ is the Heaviside step function. We calculate $K$ using the expression of Witt (1990). Using  $S$ and $p(K)dK$, the statement at the end of Sec. \ref{magmaps}, relating the microlens mass function to the magnification pattern can be written as:
\begin{equation}
\label{psf}
S(\Delta y)=\frac{1}{N_S}\int mp(m)S_{1M_{\odot}}\left(\frac{\Delta y}{\sqrt{m}}\right)dm
\end{equation}
and
\begin{equation}
\label{pff}
p(K)=\frac{1}{N_K}\int \sqrt{m}p(m)p_{1M_{\odot}}\left(K m^{-\frac{1}{4}}\right)dm.
\end{equation}
Here $S_{1M_{\odot}}$ and $p_{1M_{\odot}}$ are the structure function and distribution of flux factors for a population of 1$M_{\odot}$ microlenses. $p(m)dm$ is the microlens mass function and $N_S$ and $N_K$ are normalising constants such that:
\begin{equation}
N_S=\int mp(m)dm
\end{equation}
and
\begin{equation}
N_K=\int \sqrt{m}p(m)dm.
\end{equation}
Eqn. \ref{pff} contains an extra factor of $\frac{1}{\sqrt{m}}$ since the length of caustic per cm$^2$ increases with $\frac{1}{\sqrt{m}}$. For comparison, the assumption that light-curve statistics only on the mean may be written as: 
\begin{equation}
\label{sinsf}
S(\Delta y)=S_{1M_{\odot}}\left(\frac{\Delta y}{\sqrt{\langle m\rangle}}\right),
\end{equation}
and
\begin{equation}
\label{sinpff}
p(K)=p_{1M_{\odot}}\left(K \langle m\rangle^{-\frac{1}{4}}\right)
\end{equation}
respectively. 

In the limits of small mass ranges ($m_{1}\rightarrow m_{2}$) or where the dominant fraction of optical depth is present in microlenses of a single mass $m_o$ (ie. $mp(m)\rightarrow m_o \delta(m_o)$), Eqns. \ref{psf} and \ref{pff} reduce to Eqns \ref{sinsf} and \ref{sinpff}. In addition Eqns. \ref{psf} and \ref{pff} are correct in the limit of low optical depth. If Eqns. \ref{psf} and \ref{pff} describe the relationship between the magnification map and the microlens mass function, then these and similar expressions offer the means to measure the mass function from microlensing light-curve statistics. In the next section we explore this idea using model microlensing light-curves.

\subsection{the microlensing models}
\label{model_invert}

\begin{figure*}
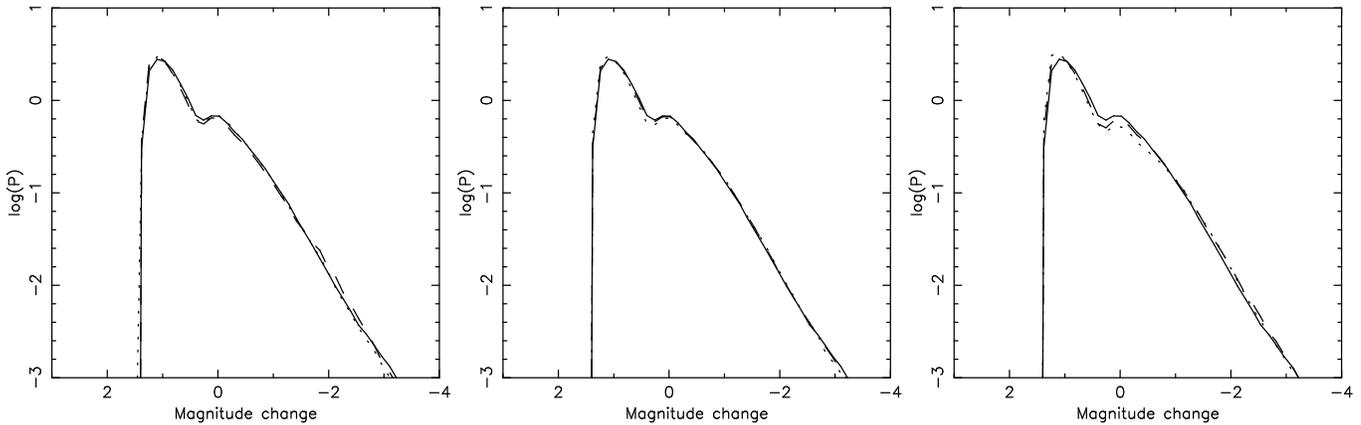

\vspace{65mm}
\includegraphics{fig4a.ps}
\includegraphics{fig4b.ps}
\includegraphics{fig4c.ps}
\caption{\label{magdist} Magnification distributions for models with $\frac{m_{2}}{m_{1}}=1$ (solid lines), $\frac{m_{2}}{m_{1}}=10$ (dashed lines) and $\frac{m_{2}}{m_{1}}=100$ (dotted lines). Left: Distributions for mass function $p_{Ndense}$, Centre: Distributions for mass function $p_{Salpeter}$, Right: Distributions for mass function $p_{Odepth}$.}
\end{figure*}

\begin{table}
\begin{center}
\caption{\label{tab3}Values for the number of stars used, and the average magnification obtained from the light-curve simulations. The theoretical mean is $\mu_{av}=2.78$.}
\begin{tabular}{|c|c|c|c|}
\hline
                                  &  \multicolumn{3}{c}{Mass function}  \\
   $\frac{m_{2}}{m_1}$   &  $p_{Ndense}$   &  $p_{Salpeter}$  & $p_{Odepth}$    \\\hline
          1              & 1644   (2.73) & 1644   (2.73)  & 1644  (2.73)\\      
        10               & 1833   (2.72) & 1791   (2.64)  & 2153  (2.70)\\
        100              & 1907   (2.67) & 2379   (2.70)  & 5622  (2.57)  \\\hline
\end{tabular}
\end{center}
\end{table}

We have investigated structure functions and distributions of flux factors produced by models with $m =1M_{\odot}$ for mass function $p_{Single}$ and by models with $\langle m\rangle =1M_{\odot}$ for mass functions $p_{Ndense}$, $p_{Salpeter}$ and $p_{Odepth}$ with $\frac{m_{2}}{m_{1}}=10$ and $\frac{m_{2}}{m_{1}}=100$. For each model, 500 point source light-curves were produced (one per random starfield), each with a length of 60 ER$_{1M_{\odot}}$ using the contouring method of Lewis et al. (1993) and Witt (1993). The number of stars in each model was determined from the description of Lewis \& Irwin (1995) (using the formalism of Katz, Balbus \& Paczynski (1986)). Table \ref{tab3} shows the number of stars and the mean magnification for each model together with the theoretical magnification. The resulting magnification distributions are shown in Fig. \ref{magdist}. Note that we confirm the finding of Lewis \& Irwin (1995) that the magnification distribution is independent of mass function.

\subsection{the structure function}
\label{sf}

\begin{figure*}
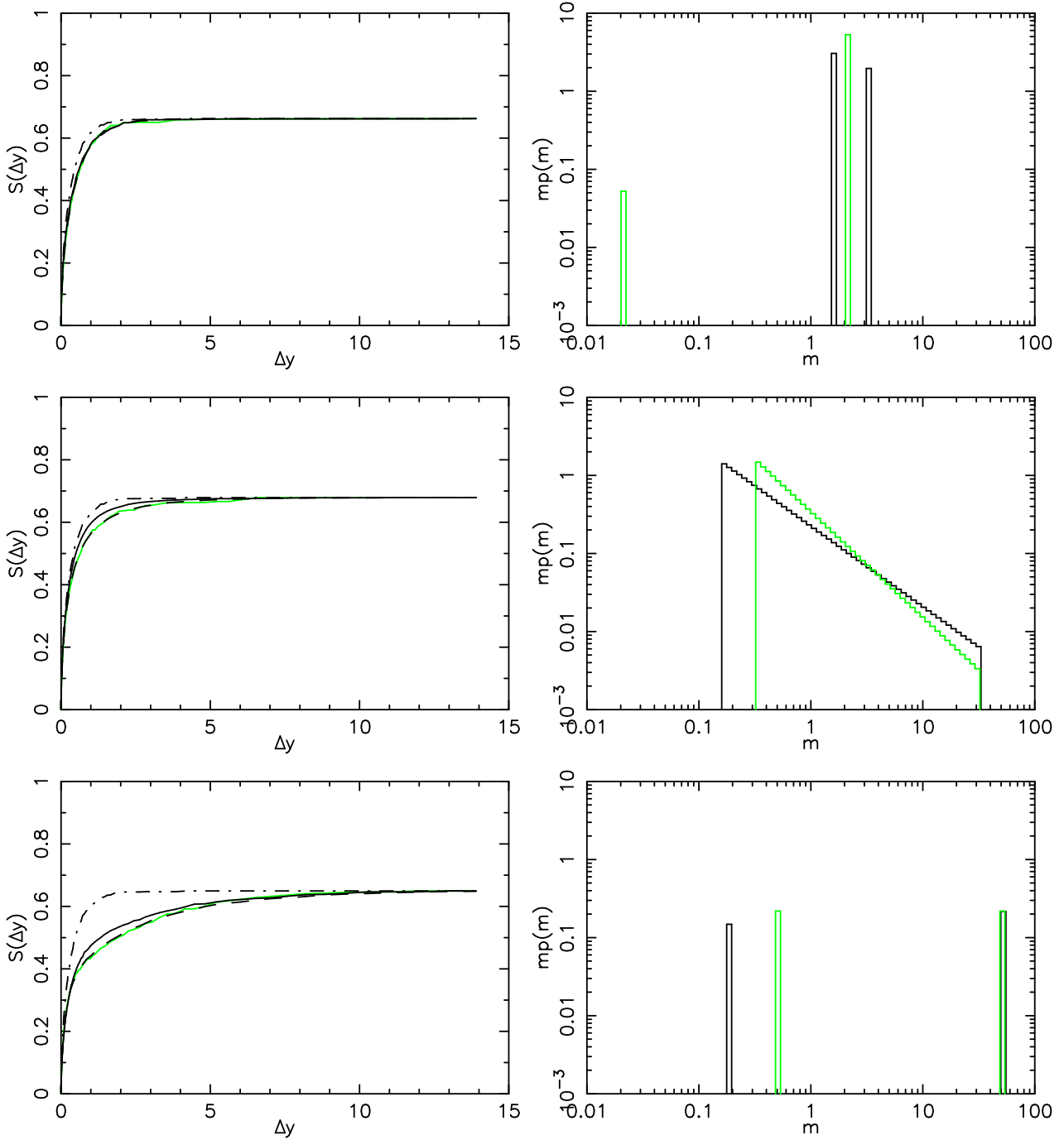

\vspace{190mm}
\includegraphics{fig5a.ps}
\includegraphics{fig5b.ps}
\includegraphics{fig5c.ps}
\caption{\label{sf} Left: The microlensing structure function. The light, dark and dashed lines show the calculated and predicted structure functions and the best fit respectively. The curve corresponding to $p_{Single}$ is shown for comparison (dot-dashed line). Right: The input mass function (light line) and the fitted mass function (dark line). Results are shown corresponding to mass functions $p_{Ndense}$ (top), $p_{Salpeter}$ (centre), and $p_{Odepth}$ (bottom). The bars in the upper and lower right-hand panels should be interpreted as having the same linear width.} 
\end{figure*}

\begin{table*}
\begin{center}
\caption{\label{tab4} Summary of input and fitted parameters for $p(m)$ obtained from Eqn. \ref{psf}, as well as the $\chi^2$ for the fit.}
\begin{tabular}{|c|c|c|c|c|c|c|c|c|}
\hline
                &\multicolumn{3}{c}{true parameters}  & \multicolumn{3}{c}{fitted parameters} &      &     \\
Mass Function   &  $\frac{m_{2}}{m_{1}}$     &  $\langle m\rangle$  &  $f\,\,/\,\,\alpha$  &  $\frac{m_{2}}{m_{1}}$ & $\langle m\rangle$ & $f\,\,/\,\,\alpha$  &  $\chi^2$ (128 points) \\\hline
      $p_{Ndense}$       &        10               &  1.0          &  0.5        &   5.15               &   1.13         &   0.17           &  $7.2\times 10^{-4}$ \\ 
      $p_{Ndense}$      &        100              &   1.0           &  0.5        &  2.05               &   1.81         &    0.37         &   $1.2\times 10^{-3}$ \\
      $p_{Salpeter}$       &        10               & 1.0          &  -2.35      &  576                 &   1.00         &    -3.06         &  $7.0\times 10^{-4}$ \\ 
      $p_{Salpeter}$      &        100              & 1.0           &  -2.35       & 186                 &   0.76         &    -2.03        & $1.0\times 10^{-3}$ \\
      $p_{Odepth}$       &        10               &  1.0           &   0.091     & 29.5                 &  0.48          &   0.03        & $8.4\times 10^{-4}$ \\ 
      $p_{Odepth}$      &        100              &   1.0           &   0.010      & 281                 &  0.41          &   0.003       & $2.5\times 10^{-3}$ \\\hline
\end{tabular}
\end{center}
\end{table*}

Eqn. \ref{psf} suggests that the structure function has a shape that reflects the microlens mass distribution. Small masses produce more rapid variability and therefore a faster rise in the structure function at small $\Delta y$. Larger masses cause the asymptotic behaviour of the structure function at large $\Delta y$ to slow. Fig. \ref{sf} shows structure functions corresponding to microlens mass functions of the form $p_{Ndense}$, $p_{Salpeter}$, $p_{Odepth}$, with $\frac{m_{2}}{m_{1}}=100$. These demonstrate the effect of a varying mass function on the structure function of a microlensed light-curve, as well as the applicability of Eqn \ref{psf}. In each case 4 curves are shown in the left-hand figure. The solid light and solid dark lines correspond to structure functions computed directly from models and from Eqn. \ref{psf} respectively. For comparison, the structure function corresponding to the single mass microlensing model ($p_{Single}$) is shown by the dot-dashed line. Fig. \ref{sf} shows that Eqn.~\ref{psf} provides a significantly better approximation to the microlensing statistics than Eqn. \ref{sinsf}. To quantify the success of the approximation we have fitted for the three free parameters of $p(m)$ in Eqn. \ref{psf}, ie. $\langle m \rangle$, $\frac{m_{2}}{m_{1}}$ and $p(m_{1})$ or the index $\alpha$ (for fits corresponding input mass functions $p_{Ndense}$ and $p_{Odepth}$ or $p_{Salpeter}$ respectively). The input mass weighted number densities (those used to compute the model light-curves) are shown as the light histograms in the right hand panels of Figs. \ref{sf}. In each case, the fitted $mp(m)$ is overlayed (dark histogram). The fitting procedure, using Eqn. \ref{psf} is far more successful for mass functions which provide significant fractions of optical depth over a large mass range. This is because the dimensions of the magnification pattern only scale with $\sqrt{m}$ (so that very different masses are needed to effect the scale of the magnification map), and also because (from Eqn. \ref{psf}) the mass function contributes to the different scales ($\sqrt{m}$) of the caustic network in proportion to the optical depth associated with mass $m$ (ie. $\kappa(m)dm$). The input and fitted parameters for $p(m)$ with $\frac{m_{2}}{m_{1}}=10,100$ are summarised in Tab. \ref{tab4}.

Fig. \ref{sf} shows excellent agreement between the directly calculated, predicted and fitted structure functions in the case of $p_{Ndense}$. However the mass function corresponding to the best fit is not in agreement with the input mass function. The reason is that the mass weighted number density for $p_{Ndense}$ is approximately that of mass function $p_{Single}$ with $m=m_{2}$. Hence $S(\Delta y)\rightarrow S_{1M_{\odot}}(\Delta y/\sqrt{m_{2}})$. The contribution of the small mass to the shape of the structure function is limited to small $\Delta y$, and $S$ less than $S(\Delta y\rightarrow\infty)\times \frac{\kappa(m_{1})}{\kappa}$. The fit has a mean mass near $m_{2}$ with a range smaller than $\frac{m_{2}}{m_{1}}$.

Fig. \ref{sf} shows reasonable agreement between the directly calculated, predicted and fitted structure functions for $p_{Odepth}$. In addition, unlike the case of $p_{Ndense}$, the fitted mass function is also in reasonable agreement with the input mass function. The contributions to the structure function from the two mass populations are distinct. Below $\Delta y\sim 0.5ER$, $S$ is dominated by the small mass population which causes rapid variability. Above $\Delta y\sim 0.5ER$, $S$ is dominated by the larger masses responsible for the longer term variability. There is a direct correspondence between two sections of S, and the two scale behaviour evident in the corresponding magnification map (Figs. \ref{magmap1} and \ref{magmap2}). Because of the break at $\Delta y\sim 0.5$, $S$ does not have a shape that is similar to a scaled version of $S_m$. The fit for $p(m)$ is therefore much more successful.

Finally, Fig. \ref{sf} also shows reasonable agreement between the directly calculated, predicted and fitted structure functions for $p_{Salpeter}$. The shape of $S(\Delta y)$ differs from that of $S_{1M_{\odot}}(\Delta y/\sqrt{\langle m\rangle})$ because the smaller masses produce a more rapid rise to the structure function, while the larger masses cause the asymptotic behaviour to slow. The input index $\alpha$ of $p_{Salpeter}$ is approximately recovered.

Our modelling is restricted to mass ranges of $\frac{m_{2}}{m_{1}}=100$ or smaller, and the effect of the mass function is subtle in all but a couple of cases. This is because of the $\sqrt{m}$ dependencies of microlensing statistics. Mass functions with much larger mass ranges would show more significant effects. However the results of this section demonstrate that Eqn. \ref{psf} is a quantitative improvement over Eqn. \ref{sinsf}, and that it is qualitatively correct in many cases. 

\subsection{the distribution of flux factors}

\begin{figure*}
\vspace{60mm}
\includegraphics{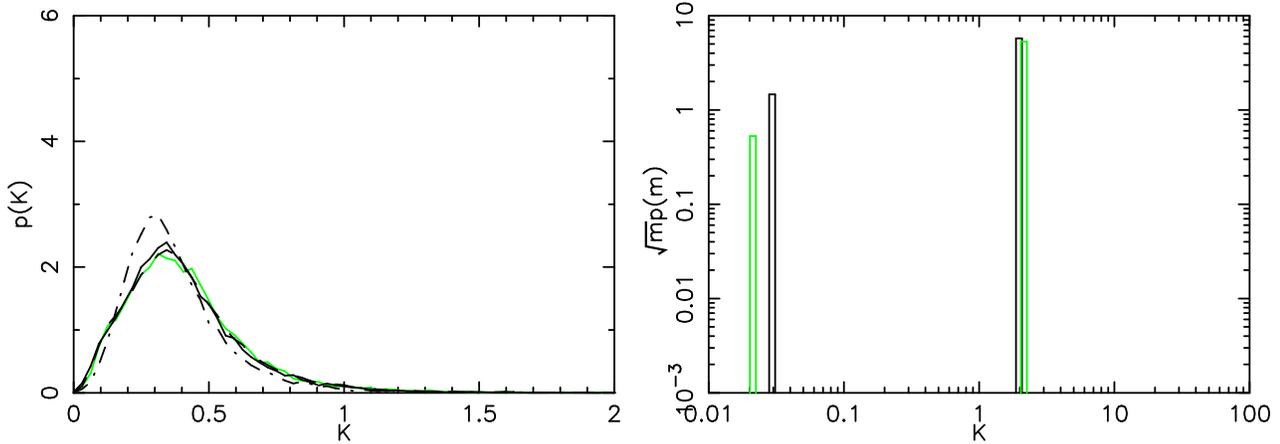}
\caption{\label{K_ND_ff} Distributions of flux factors for microlensing by a two mass population ($p_{Ndense}$). The light, dark and dashed lines show the calculated and predicted distributions and the best fit respectively. The curve corresponding to ($p_{Single}$) is shown for comparison (dot-dashed line). Right: The mass function for $p_{Ndense}$ (light line) and the fitted mass function (dark line).} 
\end{figure*}

\begin{figure*}
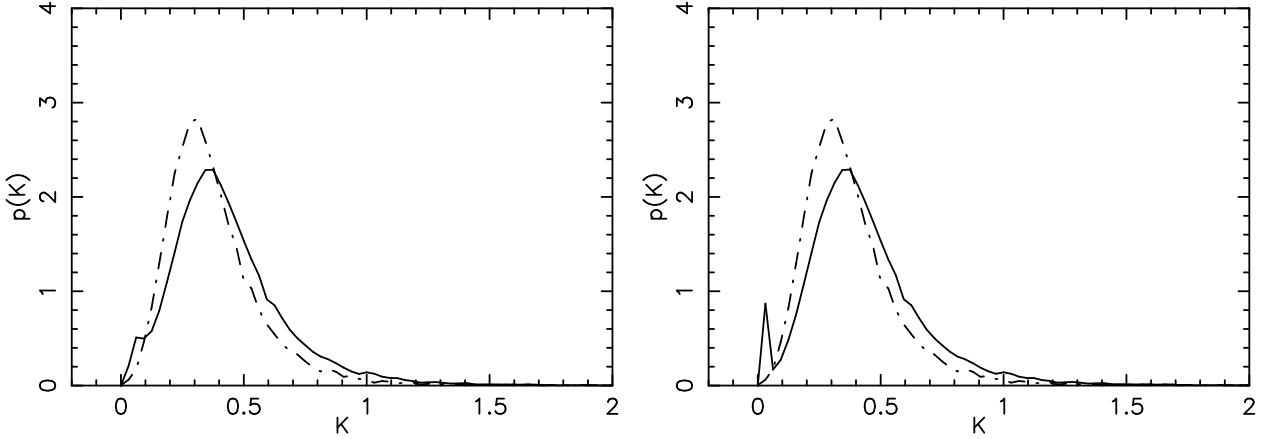

\vspace{60mm}
\includegraphics{fig7a.ps}
\includegraphics{fig7b.ps}
\caption{\label{K_ND_ff_pred} Predictions of distributions of flux factors (Eqn. \ref{pff}) for microlensing by a two mass population. Left: The distribution for mass function $p_Ndense$ with $\frac{m_{2}}{m_{1}}=1000$. Right: The distribution for a two mass population with $\frac{m_{2}}{m_{1}}=100000$, and $p(m_{1})=10p(m_{2})$. The curve corresponding to $p_{Single}$ is shown for comparison (dot-dashed line). } 
\end{figure*}

\begin{table*}
\begin{center}
\caption{\label{tab5}Summary of input and fitted parameters for $p(m)$ obtained from Eqn. \ref{pff}, as well as the $\chi^2$ for the fit.}
\begin{tabular}{|c|c|c|c|c|c|c|c|c|}
\hline
                &\multicolumn{3}{c}{true parameters}  & \multicolumn{3}{c}{fitted parameters} &      &     \\
Mass Function   &  $\frac{m_{2}}{m_{1}}$     &  $\langle m\rangle$  &  $f\,\,/\,\,\alpha$  &  $\frac{m_{2}}{m_{1}}$ & $\langle m\rangle$ & $f\,\,/\,\,\alpha$  &  $\chi^2$ (128 points) \\\hline
      $II$       &        10               &  1.0          &  0.5        &   6.30               &   0.89         &   0.11           &  $4.6\times 10^{-2}$ \\ 
      $II$      &        100              &   1.0           &  0.5        &  66.90              &   0.59         &    0.01         &   $0.1$ \\
      $III$       &        10               & 1.0          &  -2.35      &  5.16                &   0.92         &    -0.75         &  $5.6\times 10^{-2}$ \\ 
      $III$      &        100              & 1.0           &  -2.35       & 783                 &   0.99         &    -3.37        & $5.5\times 10^{-2}$ \\
      $IV$       &        10               &  1.0           &   0.091     & 5.15                 &  0.906         &   0.17        & $5.5\times 10^{-2}$ \\ 
      $IV$      &        100              &   1.0           &   0.010      & 9.53                &  0.863         &   0.10        & $9.6\times 10^{-2}$ \\\hline
\end{tabular}
\end{center}
\end{table*}

The previous section demonstrated the dependence of microlensing variability statistics on the microlens mass function. The size of a light-curve peak due to a caustic crossing is dependent on the flux factor $K$ (a function of microlens mass). In addition to the dependence of the variability, it is therefore reasonable to suppose that the distribution of peak heights should be related to the mass weighted number density, with an additional weighting of $\frac{1}{\sqrt{m}}$ to account for the higher caustic density (Eqn. \ref{pff}).

The same set of simulations and fits discussed in Sec. \ref{sf} were made for $p(K)dK$, and the results are summarised in Tab. \ref{tab5}. The extra factor of $\frac{1}{\sqrt{m}}$ in Eqn. \ref{pff}, and the dependence on $m^{\frac{1}{4}}$ rather than $\sqrt{m}$ means that $p(K)dK$ and $S(\Delta y)$ carry different information. In particular, for the case of equal optical depths in two disparate mass populations $p(K)dk\rightarrow p_m(Km_{1}^{-\frac{1}{4}})$ as $m_{2}/m_{1}$ becomes large. As a consequence, in contrast to fits using $S$, the fits for $p_{Salpeter}$ and $p_{Odepth}$ are poor. However, the factor of $\frac{1}{\sqrt{m}}$ boosts the contribution of the small masses in $p_{Ndense}$, results for which are shown in Fig. \ref{K_ND_ff} (for $\frac{m_{2}}{m_{1}}=100$). The calculated, predicted and fitted functions $p(K)dK$ show excellent agreement, and the fitted mass function is also good. The reason for the more accurate recovery of the mass function is the bump in the distribution at small $K$ corresponding to caustics related to the small masses.  

Fig. \ref{K_ND_ff_pred} shows two predicted flux factor distributions, calculated using Eqn. \ref{pff}. The left hand figure shows the distribution produced by $p_{Ndense}$, with $m_{2}=1M_{\odot}$ and $\frac{m_{2}}{m_{1}}=1000$ (row 4 of column 2 of Figs. \ref{magmap1} and \ref{magmap2}). This distribution shows a ``foot'' which is the signature of the small ($\sim$Jupiter mass) microlenses. The right hand figure shows the distribution produced by a two population mass function with $p(m_{1})=10p(m_{2})$, $m_{2}=1M_{\odot}$ and $\frac{m_{2}}{m_{1}}=100000$. The very low ($\sim$Earth) mass objects contribute a spike feature to the $p(K)$ near $K=0$.

\section{preliminary application to Q2237+0305}
\label{2237app}

\begin{figure}
\vspace{70mm}
\includegraphics{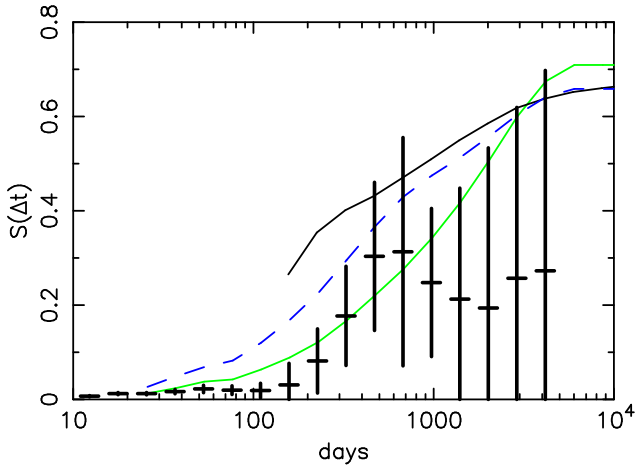}
\caption{\label{2237_SF}Binned structure function averaged over the 4 images of Q2237+0305. Superimposed are model structure functions for both a single mass model ($p_{Single}$) with $m=0.1M_{\odot}$ (light line) and a $p_{Odepth}$ model with $m_{2}=0.1M_{\odot}$ and $\frac{m_{2}}{m_{1}}=100$ (dark line). The dashed line is an approximate structure function for the $p_{Odepth}$ model computed using Eqn. \ref{psf}. The assumed effective transverse velocity was $400\,km\,sec^{-1}$, and the source size 0.13 ER$_{m_{2}}$.}
\end{figure}

Lewis et al. (1996) found that the structure function for Q2237+0305 (Huchra et al. 1985) suggests a mean mass of $0.1M_{\odot}\la \langle m\rangle\la 10M_{\odot}$. This finding remains true in light of additional light-curve data. There are currently $\approx 190$ light-curve points for each of the 4 images of Q2237+0305, spanning a baseline of $\approx 5000$ days (Schneider et al. 1988; Kent \& Falco 1989; Irwin et al. 1989; Corrigan et al. 1991; $\O$stensen et al. 1996; Wozniak et al. 2000a,b). In the past three seasons, the OGLE collaboration (Wozniak et al. 2000a,b) has obtained monitoring data at a high-rate and with good photometric accuracy. For the first time this data probes short duration fluctuations. Fig. \ref{2237_SF} presents the binned structure function, averaged over the 4 images of Q2237+0305. Quoted errors in the data have been added in quadrature for each pair, and subtracted from the square of the magnitude difference before binning. The error bars for the structure function were estimated as the variance in each bin $(\Delta t)$ divided by the squareroot of the the monitoring period over $\Delta t$ or the number of points, whichever is less. 

Superimposed on these plots are the model structure functions computed for a source size of 0.13 ER$_{0.1M_{\odot}}$ and a transverse velocity of $400\,km\,sec^{-1}$. The solid dark line shows the structure function for $p_{Odepth}$ with $m_{2}=0.1M_{\odot}$ and $\frac{m_{2}}{m_{1}}=100$. This was computed using a magnification map ($\kappa=0.4$, $\gamma=0$) of side-length 20 ER$_{1M_{\odot}}$ and composed of $500\times500$ pixels. The dashed line shows the corresponding structure function computed to smaller $\Delta t$ using Eqn. \ref{psf} and higher resolution single mass ($m$) microlensing magnification maps (side-length 10$\sqrt{\frac{m}{m_{2}}}$ER$_{1M_{\odot}}$ and composed of $1000\sqrt{\frac{m}{m_{2}}}\times1000\sqrt{\frac{m}{m_{2}}}$ pixels). The light solid line shows the structure function for $p_{Single}$ with $m=0.1M_{\odot}$ computed from the $1000\times1000$ pixel magnification maps.

In Wyithe, Webster \& Turner (2000) it was demonstrated that the source size must be smaller than $\sim$0.2 ER$_{m_{2}}$ from the large scale variability of recent seasons. The assumption of a source smaller than 0.13 ER$_{m_{2}}$ will result in more rapid initial rise of the structure function for the model with $p_{Odepth}$. Note that for this model, the source is $\sim 1$ER$_{m_{1}}$, and that the asymptotic value is slightly lower, which corresponds to a narrower magnification distribution. Both model structure functions approach their asymptotic value at around 5000$-$10000 days, reflecting the microlensing time-scale of the $0.1M_{\odot}$ microlenses. 5000 days is approximately the length of the monitoring period, and is therefore the separation where the observed structure function is least determined. However, the structure function is better determined at shorter separations. In particular if the microlens population in Q2237+0305 can be described by $p_{Odepth}$ with $m_{2}=0.1M_{\odot}$ and $\frac{m_{2}}{m_{1}}=100$ (dark solid line), then the effective sampling length for the structure function at $t\la500$ days is increased by a factor of 10, so that it becomes representative on these time-scales. However qualitative comparison between the models and the structure functions for Q2237+0305 does not show evidence for a contribution to the optical depth from very low mass microlenses. A proper treatment of this result will require models with appropriate values for $\kappa$ and $\gamma$, as well a monte-carlo analysis of structure functions modelled using sampling rates and simulated errors corresponding to those of the observations. Such an analysis will be presented in a future paper.

\section{Discussion}
\label{discussion}

The microlensing magnification pattern has a structure that reflects the mass spectrum of microlenses. In particular, variability statistics for a given, mass function are are well approximated by the average of variability statistics for single microlens mass distributions, weighted by the corresponding microlens mass and number density. Put another way, microlenses of different masses contribute to the variability statistics in proportion to $\kappa(m)dm$. In addition, High Magnification Event (HME) statistics for a particular mass function are well approximated by the average of HME statistics for single microlens mass distributions, weighted by the microlens mass ($m$), the number density, and $\frac{1}{\sqrt{m}}$, where the last factor is included to account for smaller masses producing more caustics per unit area.

If there is a similar contribution to the optical depth from masses at opposite ends of the mass range, the variability statistics are non-degenerate with the single mass case, and the mass function can be inferred by inverting an integral (Eqn. \ref{psf}). We have carried out simple parametric fits for the mass function using models with known distributions to test this idea. In practice however, more sophisticated, non parametric inversions could be used to obtain the mass function from variability data. On the other hand, since Eqn. \ref{psf} is an approximation, an inversion will most readily be achieved by directly comparing computed model statistics with data. 

 The distribution of flux factors describes the size of HME peaks. The distribution shape is sensitive to a different set of mass functions than the structure function. In particular, since smaller masses contribute more caustic length per unit area, the statistics of HMEs provide information on the mass function where the number density of objects at different ends of the mass range, rather than the optical depth is comparable. Brown dwarfs or planetary mass objects contribute to the flux factor distribution as an excess of small values. Our models assume random distributions of objects, and it is not clear what effect the spatial correlation of the large and small masses will have on the magnification map. However, in projection, and at optical depths of $\kappa\approx 0.5$, planets with orbital radii $>100$AU are not spatially correlated with stars. Microlensing of quasars by galaxies at moderate redshift may therefore shed light on the abundance of planets in a regime not accessible by other methods.

Our investigation has concentrated on what information about the distribution of microlens masses is manifest in magnification maps, and in the corresponding light-curve statistics. However, in practise obtaining sufficient quality data will be difficult. Currently, only one object, Q2237+0305, is known to exhibit significant quasar microlensing. However, the Sloan Digital Sky Survey will obtain spectra of $10^6$ galaxies (York et al. 2000), and it is estimated that a few 10s of these may have very close alignment with a background quasar, as well as a small redshift, and will therefore be efficient microlenses like Q2237+0305 (Fukugita \& Turner, Private Communication). With a large number of monitored objects, statistics of microlensed light-curves will become much better known.

 We have computed statistics for a point source and given transverse velocity, while in practice both the source size and transverse velocity will be unknown. However techniques have been developed to estimate these quantities from monitoring data (Wyithe, Webster \& Turner 1999,2000; Wyithe, Webster, Turner \& Mortlock 2000). Long light-curves extending over 10s of Einstein Radii for a 1$M_{\odot}$ star will not be available for a century or more. However the structure function effectively levels out after $\sim2$ microlens (mean) Einstein Radii. Therefore, for typical galactic proper motions, $\sim 20$ years of monitoring for multiple objects should yield an observable signal.

The detailed shape of the mass distribution is contained in the combination of a structure function and distribution of flux factors of sufficient quality. However the structure function can be used to place limits on the fraction of optical depth in small objects from the initial slope. Moreover, the time scale at which the structure function levels out yields the upper limit. The more moderate goal of probing the mass range or the existence of two distinct mass populations should therefore be achievable. Similarly, limits on the existence of small mass objects will come from the lack of observation of small light-curve peaks.

\section*{Acknowledgements}

This work was supported by NSF grant AST98-02802. JSBW acknowledges the support of an Australian Postgraduate award and a Melbourne University Overseas Research Experience Award. We would like to thank Joachim Wambsganss, Rachel Webster and Peter Schneider for helpful discussions.

\label{lastpage}

\end{document}